\documentclass[aps,twocolumn,superscriptaddress,prb]{revtex4}

\usepackage{graphicx}
\usepackage{textcomp}
\usepackage{amsmath}
\usepackage{bm}
\usepackage{subfigure}
\usepackage{color}
\graphicspath{{figures/}}

\begin{document}

\title{Enhanced metrology using preferential orientation of nitrogen-vacancy centers in diamond}

\author{L. M. Pham}
  \affiliation{School of Engineering and Applied Sciences, Harvard University, Cambridge, MA 02138, USA}
\author{N. Bar-Gill}
  \affiliation{Department of Physics, Harvard University, Cambridge MA 02138, USA}
  \affiliation{Harvard-Smithsonian Center for Astrophysics, Cambridge, MA 02138, USA}
\author{D. Le Sage}
\author{C. Belthangady}
  \affiliation{Harvard-Smithsonian Center for Astrophysics, Cambridge, MA 02138, USA}
\author{A. Stacey}
\author{M. Markham}
\author{D. J. Twitchen}
  \affiliation{Element Six Ltd., King's Ride Park, Ascot, Berkshire, SL5 8BP, UK}
\author{M. D. Lukin}
  \affiliation{Department of Physics, Harvard University, Cambridge MA 02138, USA}
\author{R. L. Walsworth}
  \affiliation{Harvard-Smithsonian Center for Astrophysics, Cambridge, MA 02138, USA}
  \affiliation{Department of Physics, Harvard University, Cambridge MA 02138, USA}

\begin{abstract}
We demonstrate preferential orientation of nitrogen-vacancy (NV) color centers along two of four possible crystallographic axes in diamonds grown by chemical vapor deposition on the $\{100\}$ face. We identify the relevant growth regime and present a possible explanation of this effect. We show that preferential orientation provides increased optical read-out contrast for NV multi-spin measurements, including enhanced AC magnetic field sensitivity, thus providing an important step towards high fidelity multi-spin-qubit quantum information processing, sensing and metrology.
\end{abstract}

\maketitle

Solid-state multi-spin-qubit systems are a promising approach to practical applications of quantum information processing, sensing and metrology. In particular, the negatively-charged nitrogen-vacancy (NV) center in diamond has a spin-triplet electronic ground-state, which can be coherently manipulated using microwave fields, optically initialized, and read out via spin-state-dependent fluorescence, with spin coherence lifetimes $>100\ \mu s$ at room temperature \cite{WrachtrupJPhys2006}. Recent demonstrations of the utility of NV centers as quantum registers and magnetic and electric field sensors have attracted great interest \cite{DuttSci2007, TaylorNatPhys2008, MazeNat2008, BalasubramanianNat2008, JiangSci2009, NeumannSci2010, StanwixPRB2010, MaurerNatPhys2010, PhamNJP2011, DoldeNatPhys2011, GrinoldsNatPhys2011, HongArXiv2012, LeSagePRB2012, MaletinskyNNano2012, BarGillNatComm2012}.

\begin{figure}[]
  \includegraphics[width=0.91\columnwidth]{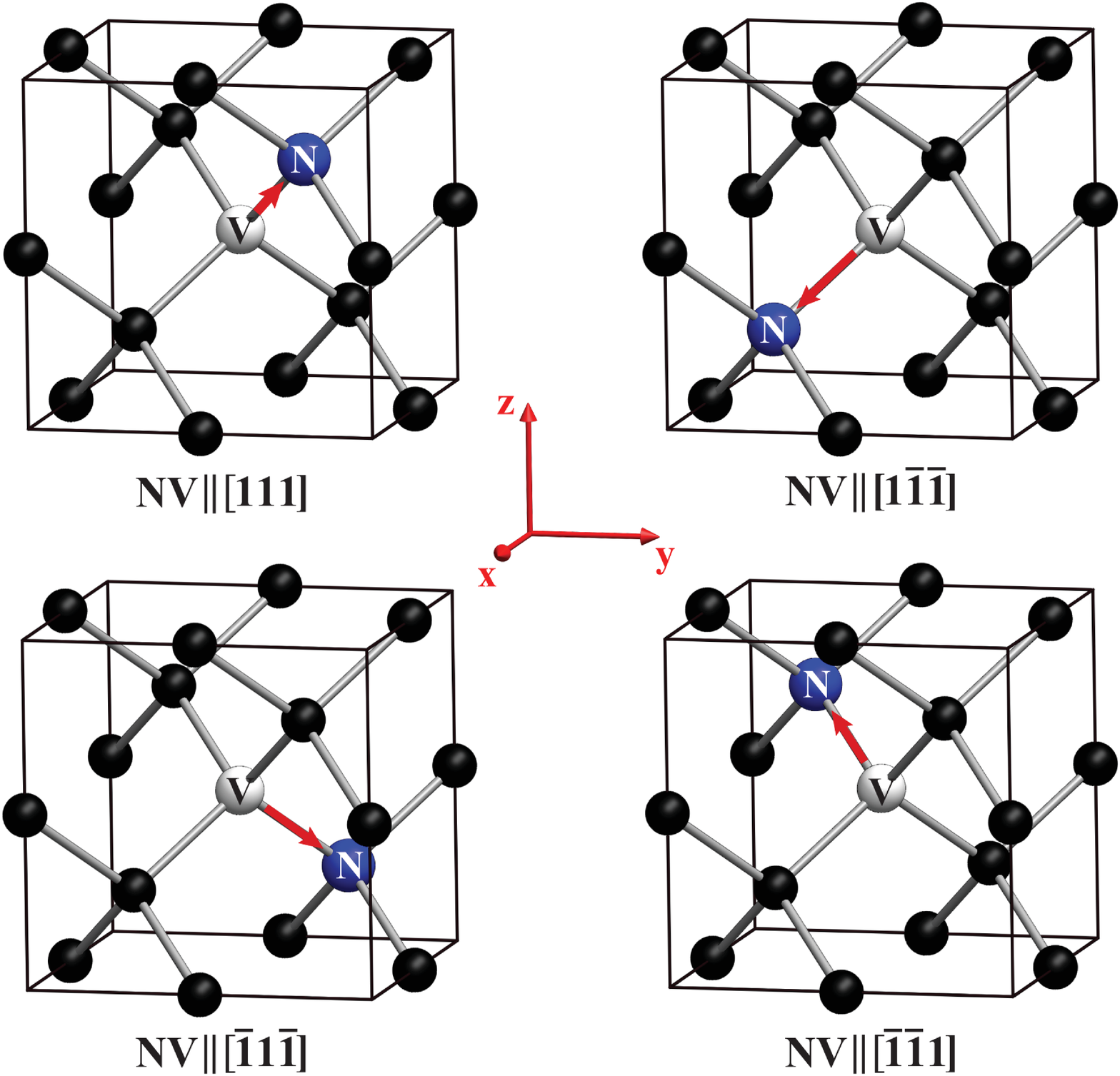}
\caption{Four possible orientations of nitrogen-vacancy (NV) color centers in diamond. Carbon atoms are depicted in black, nitrogen (N) atoms in blue, and vacancies (V) in white. NV electronic spin is indicated by red arrows.} \label{fig:intro}
\end{figure}

The NV center is composed of a substitutional nitrogen (N) impurity and a vacancy (V) on adjacent lattice sites in the diamond crystal. Due to the diamond crystal structure, NV centers can be classified by the orientation of their symmetry axes along one of four possible crystallographic axes: $[111],[1 \bar{1} \bar{1}],[\bar{1} 1 \bar{1}],$ and $[\bar{1} \bar{1} 1]$ (see Fig. \ref{fig:intro}). In most diamond samples, NV centers occupy these four orientations equally. To spectrally distinguish a single NV orientation class, a static magnetic field is applied along the relevant crystallographic axis, which maximally splits the degeneracy between $m_s = \pm 1$ sublevels for the desired NV orientation class \cite{PhamNJP2011, StanwixPRB2010}. This typically limits coherent NV spin manipulation via a resonant RF field to only one quarter of the NV population, with the rest of the NV centers contributing background fluorescence and thereby degrading read-out contrast. Thus, preferential orientation of NV centers along a subset of the four crystallographic axes would benefit multi-spin applications by both employing a greater fraction of the NV center population and reducing background fluorescence.

It was recently shown that for synthetic diamond grown via chemical vapor deposition (CVD) on $\{110\}$-oriented substrates, NV centers can be incorporated into the lattice as a unit and thereby found in only two orientations \cite{EdmondsArxiv2011}. Here we show that preferential NV orientation can also be realized for CVD diamond samples grown on $\{100\}$-oriented substrates, which are more commonly available and more compatible with increased area bulk production\cite{SilvaDRM2009}, making them more suitable for applications such as bulk magnetometry. We describe the growth regime and likely mechanism leading to preferential NV orientation.  We then present experimental demonstrations that diamond with preferential NV orientation exhibits higher contrast for optical read-out of the NV spin state and as a result provides improved AC magnetic field sensitivity.

We employed a wide-field fluorescence microscope for optically-detected electron spin resonance (ESR) measurements on ensembles of NV centers \cite{StanwixPRB2010, PhamNJP2011, BarGillNatComm2012} in two $\{100\}$-oriented CVD-grown bulk diamond samples: one with NV centers populating the four orientation classes equally (Sample A) and one with preferential NV orientation along two crystallographic axes (Sample B). The two samples have similar densities of NV centers (NV $\sim 5 \times 10^{12}\ \rm{cm}^{-3}$ for Sample A, NV $\sim 3 \times 10^{12}\ \rm{cm}^{-3}$ for Sample B) and spin coherence lifetimes ($T_2 \approx 480 \mu s$ for Sample A, $T_2 \approx 530 \mu s$ for Sample B). Sample A contains a natural isotopic abundance of $^{13}$C (1.1\%), which has nuclear spin 1/2, whereas Sample B is isotopically pure (0.01\% $^{13}$C). For each sample, NV centers within a $10\ \mu$m-wide region were optically excited by a switched 3-Watt 532-nm laser; the resulting NV spin state-dependent fluorescence ($\sim 640 - 800$ nm) was collected by a microscope objective and imaged onto a CCD array. A microwave (MW) field for NV spin state manipulation was generated by a loop antenna designed to produce a homogeneous field over the sample detection volume.

\begin{figure}[]
  \includegraphics[width=0.91\columnwidth]{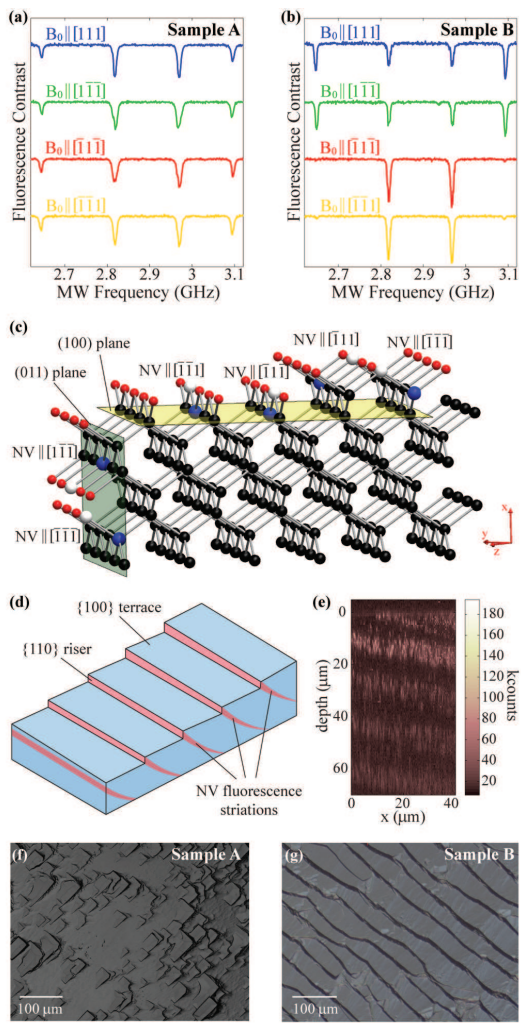}
\caption{ESR spectra with static magnetic field ($B_0 = 80$ Gauss) along each of  four diamond crystallographic axes for (a) Sample A, which exhibits no preferential NV orientation, and (b) Sample B, which exhibits a high level ($\sim 94\%$) of preferential NV orientation. (c) Atomic-level model of NV incorporation in $(100)$ step-flow growth, indicating mechanism for preferential NV occupation of only two orientations. (d) Diagram of $\{100\}$ step-flow growth surface and fluorescence striations. (e) Confocal cross-section of a sample which exhibits fluorescence striations characteristic of $\{100\}$ step-flow growth. Microscope images of the surface morphologies of (f) Sample A and (g) Sample B.} \label{fig:100}
\end{figure}

To determine the relative population of the four NV orientation classes in the two samples, we applied a static field ($B_0 = 80$ Gauss) along each of the four diamond crystallographic axes (i.e., NV axes) and measured the resulting ESR spectra. For such magnetic field configurations, four NV ESR resonances are observed: one pair of resonances corresponds to the $m_s=0$ to $m_s = \pm 1$ transitions for the class of on-axis NV centers  that are oriented parallel to $B_0$ (at 2.64 GHz and 3.10 GHz for $B_0 = 80$ Gauss); another pair of NV ESR resonances corresponds to the three classes of off-axis NV centers that are not aligned with $B_0$ (at 2.82 GHz and 2.97 GHz for $B_0 = 80$ Gauss). In Sample A, the observed amplitudes of the NV resonances in the ESR spectra were very similar for each of the four static magnetic field configurations, indicating equal population of NV centers in the four orientation classes (see Fig. 2(a)). In Sample B, however, the pair of on-axis NV ESR resonances were found to be very weak (i.e., small amplitude) for the static magnetic field configurations $B_0 \parallel [\bar{1} 1 \bar{1}]$ and $B_0 \parallel [\bar{1} \bar{1} 1]$, indicating a high fraction of NV centers with preferential orientation (see Fig. 2(b)). From the measured relative amplitudes of the two pairs of NV ESR resonances, we estimate that $\sim 94\%$ of the NV centers in Sample B are oriented along either the $[111]$ or $[1 \bar{1} \bar{1}]$ directions.

In the idealized picture of CVD growth on the $(100)$ surface  (or any of the symmetrically equivalent \{100\} faces), a substitutional nitrogen atom can be incorporated at a lattice site in two configurations: (i) the two remaining bonds above the nitrogen allow a vacancy to form an NV center with a $[\bar{1} \bar{1} 1]$ or $[\bar{1} 1 \bar{1}]$ orientation with equal probability or (ii) the two remaining bonds above the nitrogen allow a vacancy to form an NV center with a $[\bar{1} 1 1]$ or $[\bar{1} \bar{1} \bar{1}]$ orientation with equal probability. Thus in the idealized picture of CVD growth on a $\{100\}$ surface, the four possible NV orientation classes form with equal probability (see Fig. 2(c)). However, for certain conditions of substrate temperature and flow of nitrogen gas through the CVD chamber, growth on a $\{100\}$ surface may occur via a step-flow mode \cite{AchardDRM2007}. In this growth mode, the surface morphology is stepped (comprising  vertical risers and horizontal terraces), as shown in Fig. 2(d), where the risers of the steps correspond approximately to a $\{110\}$ crystallographic plane (within $5^{\circ}$). Furthermore, in the step-flow growth regime nitrogen atoms are incorporated more readily into the risers rather than the terraces \cite{MartineauGG2004}, which results in visible striations in the fluorescence image of the diamond, as shown in Fig. 2(e). Due to the preferential incorporation of nitrogen---and hence the enhanced creation of NV centers---in $\{110\}$ risers, preferential NV orientation can be realized in $\{100\}$-oriented CVD diamond via the same mechanism that allows preferential NV orientation in $\{110\}$-oriented CVD diamond \cite{EdmondsArxiv2011}.

As seen in Fig. 2(f), the surface morphology of Sample A is not stepped, indicating that it was not grown under step-flow conditions, which is consistent with the observed lack of preferential NV orientation in this sample (Fig. 2(a)). In contrast, the surface morphology of Sample B is stepped (Fig. 2(g)), indicating that it was grown under step-flow conditions, which is consistent with the observed preferential NV orientation of this sample (Fig. 2(b)). We also observed varying levels of preferential NV orientation in several additional $\{100\}$-oriented CVD samples exhibiting stepped surface morphology and fluorescence striations characteristic of step-flow growth. Further characterization of growth factors determining the level of preferential NV orientation will be the focus of future work.

We emphasize that the advantage of samples with preferential NV orientation for NV multi-spin applications arises from practical issues that typically restrict useful measurements to one NV orientation class at a time. In particular, precision coherent NV spin manipulations (e.g., magnetometry) require application of high-accuracy, resonant MW pulses. Thus, it is usually necessary to select one NV orientation class and then treat this NV class as an effective multi-spin two-level system by applying a static magnetic field to split the degeneracy between the $m_s = \pm 1$ states \cite{TaylorNatPhys2008}. It is also necessary for the static field to be well-aligned with the selected NV axis to avoid enhanced NV spin decoherence due to anisotropic hyperfine interactions with proximal $^{13}$C nuclear spin impurities \cite{StanwixPRB2010, MazePRB2008}. It is possible to avoid this limitation by using an isotopically pure $^{12}$C sample with the static field aligned such that the different NV classes are energetically degenerate. However in this case, there are two issues: an equal spin Rabi frequency is needed for all four NV orientation classes in order to drive all NV spins coherently, which is technically quite challenging; and for magnetometry, vector information about the magnetic field being sensed is lost. Therefore, given these constraints, it is generally desirable to manipulate a single NV class, in which case fluorescence from other NV orientations contributes to the background signal. Hence in samples with NVs oriented only along two axes, we expect a factor of $\sim 2$ improvement in read-out contrast compared to standard samples with equal population of all NV orientations. Note also that since the NV optical transition electric dipole moment lies in the plane perpendicular to the NV axis, a proper choice of the optical polarization of the excitation laser beam can further increase the NV read-out contrast \cite{EpsteinNat2005}, which translates into additional improvement in metrology sensitivity and quantum processing fidelity.

\begin{figure}[]
  \includegraphics[width=0.91\columnwidth]{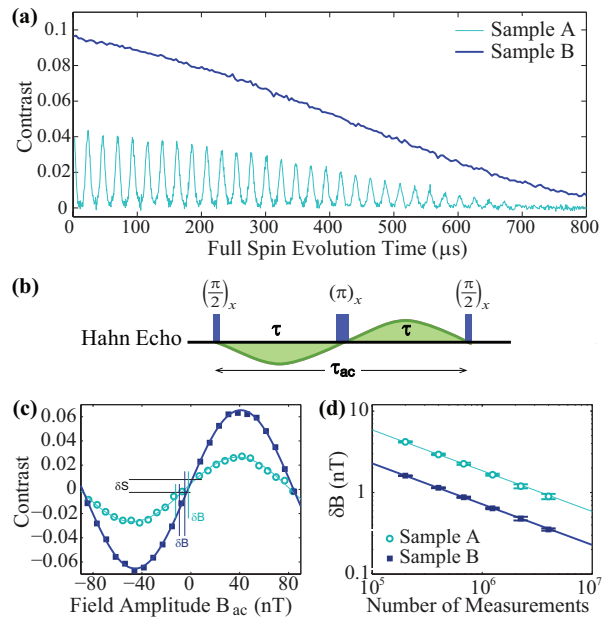}
\caption{Comparison for NV non-oriented Sample A and preferentially-oriented Sample B of (a) NV spin coherence decay measured using a (b) Hahn echo pulse sequence (timing of AC magnetic field to be measured is shown in green); (c) NV magnetometry signal (i.e., fluorescence contrast) as a function of AC magnetic field amplitude acquired using same method as in \citep{PhamNJP2011}; and (d) AC magnetic field sensitivity as a function of measurement time. Spin-state-dependent optical read-out contrast from Sample B is $\gtrsim 2$ times larger than from Sample A, leading to similarly improved magnetic field sensitivity.} \label{fig:contrast}
\end{figure}

In Fig. 3(a) we present NV spin coherence measurements that demonstrate the expected factor of $\gtrsim 2$ improvement in NV spin-state-dependent optical read-out contrast for Sample B  (preferential NV orientation along two axes) relative to Sample A (equal populations of all NV orientations). We compare the NV spin coherence decay of the two samples, measured using a Hahn Echo pulse sequence \cite{HahnPhysRev1950} as follows: all NVs are optically initialized to the $m_s=0$ state \cite{ChildressSci2006}; NVs oriented along the static magnetic field direction are then subjected to a $\pi/2$ - $\tau$ - $\pi$ - $\tau$ - $\pi/2$ pulse sequence; and finally all NVs are read out optically. By varying the duration between pulses $\tau$, the decay of NV spin coherence as a function of time is measured. Note that Sample A exhibits collapses and revivals in the measured NV spin coherence, which result from the Larmor precession of $^{13}$C impurities in this sample \cite{ChildressSci2006}. This effect does not alter the decoherence envelope, which is similar for both samples.

The improvement in optical read-out contrast for samples with preferential NV orientation is enabling for high fidelity multi-spin metrology. For example, in a standard AC magnetometry measurement utilizing a Hahn Echo control pulse sequence, an AC magnetic field (collinear with the static magnetic field $B_0$) induces a net phase accumulation in the coherence of NV spins that have their axis oriented along $B_0$, which is then translated into a measured NV fluorescence (i.e., magnetometry) signal \citep{TaylorNatPhys2008, MazeNat2008, PhamNJP2011}. The accumulated phase scales linearly with the amplitude of the magnetic field and is maximized when the full time of the Hahn Echo sequence is equivalent to the period of the AC magnetic field ($\tau_{ac}$) and the control pulses coincide with nodes in the AC magnetic field (Fig. 3(b)). Under these conditions, optimum AC magnetic field sensitivity is achieved, as given approximately by \cite{TaylorNatPhys2008}:
\begin{equation}\label{eq:HEsens}
\eta \approx \frac{\pi \hbar}{2 g \mu_B} \frac{1}{C(T) \sqrt{\tau_{ac} n_{NV}}}
\end{equation}
Here $C$ is a parameter that depends on the optical collection efficiency and measurement contrast, and $n_{NV}$ is the number of NV spins contributing to the measurement. In particular, $C$ scales linearly with the measurement contrast, which is modified by NV spin decoherence over the measurement time $T \approx \tau_{ac}$.

In Fig. 3(c) we plot the measured NV magnetometry signal (i.e., fluorescence contrast) as a function of applied AC magnetic field amplitude for both the preferentially oriented Sample B and the standard (non-oriented) Sample A, using the Hahn Echo sequence described above with an AC magnetic field of frequency $f_{ac} = 3.08$ kHz. The enhanced read-out contrast from the preferentially oriented sample improves the magnetic field sensitivity by a factor $\gtrsim 2$. As shown in Fig. 3(d), we find that the AC magnetic field sensitivity for the preferentially oriented Sample B is $18.4 \pm 1.3 \rm{nT/\sqrt{Hz}}$, whereas the sensitivity for the standard Sample A is $47.1 \pm 2.5 \rm{nT/\sqrt{Hz}}$. The samples we used in these demonstration measurements are similar in all parameters except for the population of the different NV classes, and therefore the better sensitivity of Sample B can be attributed to the better read-out contrast provided by preferential NV orientation.

In summary, we demonstrated that CVD diamond grown on the $\{100\}$ crystallographic face can yield NV color centers that are preferentially oriented along two of the four possible crystallographic axes. We attribute this effect to the creation of terraces and risers in the step-flow growth morphology, such that NVs are incorporated mainly into the risers, which correspond approximately to $\{110\}$ surfaces and therefore accommodate only two NV orientations. We showed that this preferential NV orientation increases the optical read-out contrast of the NV spin state by about a factor of two, and enables a similar enhancement in AC magnetometry sensitivity. These findings will allow the design and growth of optimized diamond samples with tailored NV orientations, paving the way for high-fidelity multi-spin-qubit quantum information processing, sensing and metrology.

\begin{acknowledgments}
This work was supported by the NSF and DARPA (QuEST and QuASAR programs). We gratefully acknowledge helpful technical discussions with Keigo Arai, David Glenn and Huiliang Zhang.
\end{acknowledgments}

\bibliography{nvpreforientnodoi}

\end{document}